\documentclass[12pt,twoside]{article}
\usepackage{xrb2007}
\usepackage{multicol}
\newbox\grsign \setbox\grsign=\hbox{$>$} \newdimen\grdimen \grdimen=\ht\grsign
\newbox\simlessbox \newbox\simgreatbox \newbox\simpropbox
\setbox\simgreatbox=\hbox{\raise.5ex\hbox{$>$}\llap
   {\lower.5ex\hbox{$\sim$}}}\ht1=\grdimen\dp1=0pt
\setbox\simlessbox=\hbox{\raise.5ex\hbox{$<$}\llap
   {\lower.5ex\hbox{$\sim$}}}\ht2=\grdimen\dp2=0pt
\setbox\simpropbox=\hbox{\raise.5ex\hbox{$\propto$}\llap
   {\lower.5ex\hbox{$\sim$}}}\ht2=\grdimen\dp2=0pt
\def\simgreat{\mathrel{\copy\simgreatbox}}
\def\simless{\mathrel{\copy\simlessbox}}

\begin{document}

\session{Obscured XRBs and INTEGRAL Sources}

\shortauthor{Walter, R. \& Zurita, J.}
\shorttitle{Probing Clumpy Stellar Wind in SFXTs}

\title{Probing Clumpy Stellar Winds in SFXTs}
\vspace{-0.6cm}
\author{R.\ Walter}
\affil{Observatoire de Gen\`eve, Universit\'e de Gen\`eve, Chemin des Maillettes 51, CH-1290 Sauverny; INTEGRAL Science Data Centre, Chemin d'Ecogia 16, CH-1290 Versoix}
\author{J.\ Zurita-Heras}
\affil{Laboratoire AIM, CEA/DSM-CNRS-Universit\'e Paris Diderot, DAPNIA/SAP,\\ F-91191 Gif-sur-Yvette}
\begin{abstract}
Quantitative constraints on the wind clumping of massive stars can be obtained from the study of the hard X-ray variability of SFXTs. In these systems, a large fraction of the hard X-ray emission is emitted in the form of flares with typical duration of 3 ksec, frequency of 7 days and luminosity of $10^{36}$ ergs/s. Such flares are most probably emitted by the interaction of a compact object orbiting at $\sim10~R_*$ with wind clumps ($10^{22-23}$ g). The density ratio between the clumps and the inter-clump medium is  $10^{2-4}$. 
The parameters of the clumps and of the inter-clump medium are in good agreement with macro-clumping scenario and line-driven instability simulations. 
\end{abstract}
\vspace{-0.6cm}
\section{Introduction}
Stellar winds have profound implications for the evolution of massive stars, on the chemical evolution of the Universe and as a source of energy and momentum in the interstellar medium. 
Indirect measures of the structure of massive-star winds are possible in X-ray binaries through the analysis of the interaction between the compact companion and the stellar wind. 
In this  report we summarize the constraints obtained on wind clumping in HMXBs using the hard X-ray variability observed by the IBIS/ISGRI instrument 
on board INTEGRAL \citep{walter:winkler03AA}. 
Classical wind-fed, Roche-lobe underflow, super-giant HMXB (sgHMXB) are made of a compact object orbiting within a few (1.5 to 2.5) stellar radii from a super-giant companion. Recently INTEGRAL almost tripled the number of sgHMXB systems known in the Galaxy and revealed a much more complex picture with two additional families of sources:
(1) the highly-absorbed systems which have orbital and spin periods similar to those of classical  sgHMXB but much higher absorbing column densities on average \citep{walter:walter2006} and (2) the fast transient systems which are characterized by fast outbursts and by a very low quiescent luminosity \citep{walter:Sguera2006,walter:Negueruela2007}.


Several sources have now been proposed as candidate super-giant fast X-ray transient based on their hard X-ray variability characteristics, and, for a subset of them, optical counterpart spectral type. Contrasting statements have however been made on specific sources for what concerns their persistent or transient nature. In the frame of the current study \citep{walter:WalterZurita2007} we have considered all SFXT candidates together with several persistent and absorbed super-giant HMXB for comparison. 

We analyzed the available INTEGRAL data for 12 candidate SFXT 
that have large variability factors and compared them with the classical and absorbed sgHMXB systems that have a typical variability factor $\simless20$.
The sources have been separated in two categories. The SFXT include systems featuring hard X-ray variability by a factor $\simgreat 100$. ``Intermediate'' systems are candidate SFXT with smaller variability factors that could be compared with those of classical systems. 

\vspace{-0.3cm}
\section{Hard X-ray Flares and Clumpy Winds}

The average count rate observed during SFXT flares lies between 3 and 60 ct/s which translates to hard X-ray luminosities of $(0.2-4)\times 10^{36}~\rm{erg/s}$. Such luminosities are not exceptional for sgHMXB but very significantly larger than the typical X-ray luminosity of single massive stars of $10^{30-33}~\rm{erg/s}$ at soft X-rays \citep{walter:Cassinelli1981}. 
As the sources are flaring at most once per day, their average hard X-ray luminosity is very low.
It is therefore very unlikely that those systems have an average orbital radius lower than $10^{13}~\rm{cm}$, i.e. $\sim 10~R_*$. One expects orbital periods larger than 15 days and underflow Roche lobe systems (note that no orbital period has yet been derived in any of these systems). 

The interaction of a compact object with a dense clump formed in the wind of a massive companion leads to increased accretion rate and hard X-ray emission. 
The free-fall time from the accretion radius $R_a = 2\times 10^{10}~ \rm{cm}$ towards the compact object is of the order of $(2-3)\times10^2~\rm{sec}$. 
The infall is mostly radial (down to the Compton radius) and proceeds at the Bondi-Hoyle accretion rate. 
With a duration of $t_{fl}=2-10$ ksec, the observed short hard X-ray flares are significantly longer than the free-fall time. The flare duration is therefore very probably linked with the thickness of the clumps which,  for a clump radial velocity $V_{cl}=10^8 ~\rm{cm/s}$, is $h_{cl} = V_{cl} \times t_{fl} \sim (2-10) \times 10^{11}~\rm{cm}$.

The average hard X-ray luminosity resulting from an interaction between the compact object and the clump can be evaluated as $L_X  = \epsilon~M_{acc}c^2/t_{fl}$ (where $\epsilon\sim0.1$) and the mass of a clump can then be estimated as
$ M_{cl} = ~ (R_{cl}/R_{a})^2 ~M_{acc}= (R_{cl}/R_{a})^2~L_X~t_{fl}/(\epsilon~ c^2) $
where $R_{cl}$ is the radius of the clump perpendicular to the radial distance.
In the case of a spherical clump,
$M_{cl} = 
\left(\frac{L_X}{10^{36}~\rm{erg/s}}\right) \left(\frac{t_{fl}}{3~\rm{ks}}\right)^3 
~7.5\times 10^{21} ~\rm{g}.$

If $\dot{N}$ is the rate of clumps emitted by the star, the observed hard X-ray flare rate is given by $T^{-1} = \dot{N}(R_{cl}^2/4R_{orb}^2).$
The rate of mass-loss in the form of wind clumps can then be estimated as
$\dot{M}_{cl}  =
\left(\frac{10\rm{d}}{T}\frac{L_X}{10^{36}\rm{erg/s}}\frac{t_{fl}}{3\rm{ks}}\right)\left(\frac{R_{orb}}{10^{13}\rm{cm}}\right)^2 ~3\times 10^{-6}~\rm{M_{\odot}/y}.$
For a $\beta=1$ velocity law and spherical clumps, the number of clumps located between $1.05R_*$ and $R_{orb}$  can be evaluated as  
$N=
\left(\frac{10~\rm{d}}{T}\right)\left(\frac{3~\rm{ks}}{t_{fl}}\right)^2\left( \frac{R_{orb}}{10^{13}~\rm{cm}}\right)^3~3.8\times 10^3$. 
Assuming spherical clumps, the clump density at the orbital radius is $\rho_{cl}=\left(\frac{L_X}{10^{36}~\rm{erg/s}}\right) ~7\times 10^{-14} ~\rm{g~cm}^{-3}$ and the corresponding homogeneous wind density is $\rho_h=\dot{M}_{cl}/(4\pi~R_{orb}^2~V_{cl})=
\left(\frac{10~\rm{d}}{T}\frac{L_X}{10^{36}~\rm{erg/s}}\frac{t_{fl}}{3~\rm{ks}}\right)
~1.5\times 10^{-15}~\rm{g~cm}^{-3}$. The clump volume filling factor at the orbital radius is $
f_V = \frac{\rho_h}{\rho_{cl}} = 
\left(\frac{10~\rm{d}}{T}\frac{t_{fl}}{3~\rm{ks}}\right)
~0.02$ and the corresponding porosity length is
$h=\frac{R_{cl}}{f_V}=
\left(\frac{T}{10~\rm{d}}\right)
~15\times 10^{12} ~\rm{cm}$.
If the density of a clump decreases with radius as $r^{-2\beta}$ and its mass remains constant, the averaged homogeneous wind density within $R_{obs}$ is  $\overline{\rho_{h}}=N M_{cl}/(\frac{4}{3}\pi 
R_{orb}^3
) = 
\left(\frac{10~\rm{d}}{T}\frac{L_X}{10^{36}~\rm{erg/s}}\frac{t_{fl}}{3~\rm{ks}}\right)
~7\times 10^{-15} ~\rm{g~cm}^{-3}$ and the average clump volume filling factor and porosity length could be estimated as 0.1 and $3\times10^{12} ~\rm{cm}$, respectively.

The variety of $t_{fl}$, $T$ and $F_{fl}$ that are observed probably reflects a range of  clump parameters and orbital radii. Several of the average clump parameters estimated above, in particular the clump density, filling factor and  porosity length do not depend on the orbital radius, which is unknown, and only slowly depend on the observed quantities.

\section{Discussion}

The average clump parameters derived above match the macro-clumping scenario of \cite{walter:OskinovaHamannFeldmeier2007}
to reconcile clumping and mass-loss rates. 
The number of clumps derived above is also comparable to evaluations by \cite{walter:Lepine1999, walter:OskinovaFeldmeierHamann2006}. The volume filling factor, porosity length and the clump mass-loss rate are also similar to those derived by \cite{walter:Bouret2005} from the study of ultraviolet and optical line profiles in two super-giant stars.


The variation of the observed X-ray flux between flares and quiescence provides in principle a direct measure of the density contrast between the wind clumps and the inter-clump medium. 
Density contrasts of $>10^{2-4}$ and 15--50 have been observed in SFXT and ``Intermediate'' sources, respectively. The density contrast is larger in SFXT than in ``Intermediate'' and, of course, classical systems. Density contrasts are probably stronger when clumping is very effective. 

Numerical simulations of the line driven instability \citep{walter:Runacres2005} predict density contrasts as large as $10^{3-5}$ in the wind up to large radii. At a distance of $10~R_*$, the simulated density can vary between $10^{-18}$ and $10^{-13}~\rm{g~cm^{-3}}$ and the separation between the density peaks are of the order of  $R_*$. These characteristics are comparable to the values we have derived.

Classical sgHMXB are characterized by small orbital radii $R_{orb}=(1.5-2.5)~R_*$, and by flux variability of a factor $\simless 10$. Such variabilities were modeled in terms of wind inhomogeneities largely triggered by the hydrodynamic and photo-ionisation effects of the accreting object on the companion and inner stellar wind \citep{walter:blondin91, walter:blondin94}. At small orbital radii, the companion is close to fill its Roche lobe, which triggers tidal streams. In addition the X-ray source ionizes the wind acceleration zone, prevents wind acceleration and generates slower velocities, denser winds, larger accretion radius and finally larger X-ray luminosities. Whether or not the stellar wind is intrinsically clumpy at low radius, the effect of the compact object on the wind is expected to be important.

The main difference between SFXT and classical sgHMXB could therefore be their orbital radius. At very low orbital radius $(<1.5~R_*)$ tidal accretion will take place through an accretion disk and the system will soon evolve to a common envelope stage. At low orbital radius $(\sim 2~R_*)$ the wind will be perturbed in any case and efficient wind accretion will lead to copious and persistent X-ray emission $(10^{36-37}~\rm{erg/s})$. At larger orbital radius $(\sim 10~R_*)$ and if the wind is clumpy, the SFXT behavior is expected as described above. If the wind clumps do not form for any reason, the average accretion rate will remain too low and the sources will remain mostly undetected by the current hard X-ray survey instruments.


\begin{thebibliography}{}
\bibitem[Blondin et al. 1991]{walter:blondin91}
Blondin, J., et al., 1991, ApJ, 371, 684
\bibitem[Blondin, 1994]{walter:blondin94}
Blondin, J., 1984, ApJ, 435, 756
\bibitem[Bouret et al. (2005)]{walter:Bouret2005}
Bouret, J.-C., et al., 2005, ApJ, 438, 301
\bibitem[Cassinelli et al. 1981]{walter:Cassinelli1981}
Cassinelli, J.-P, et al., 1981, ApJ, 250, 677
\bibitem[Illarionov and Beloborodov, 2001]{walter:Illarionov2001}
Illarionov, A., et al., 2001, MNRAS, 323, 159
\bibitem[Lepine and Moffat (1999)]{walter:Lepine1999}
Lepine, S., Moffat, A., 1999, ApJ, 514, 909
\bibitem[Negueruela et al. 2007]{walter:Negueruela2007}
Negueruela, I, et al., 2007, arXiv:0704.3224v2
\bibitem[Oskinova et al. (2006)]{walter:OskinovaFeldmeierHamann2006}
Oskinova, L.~M., et al. 2006, MNRAS, 372, 313
\bibitem[Oskinova et al. (2007)]{walter:OskinovaHamannFeldmeier2007}
Oskinova, L.~M., et al., A., 2007, arXiv:0704.2390
\bibitem[Runacres and Owocki, 2005]{walter:Runacres2005}
Runacres, M., Owocki, S., 2005, A\&A, 429, 323
\bibitem[Sguera et al. 2006]{walter:Sguera2006}
Sguera, V., et al., 2006, ApJ, 646, 452
\bibitem[Walter et al. 2006]{walter:walter2006}
Walter, R., et al., 2006, A\&A, 453, 133
\bibitem[Walter and Zurita (2007)]{walter:WalterZurita2007}
Walter, R., Zurita-Heras, J., 2007, A\&A, 476, 335
\bibitem[Winkler et al. 2003]{walter:winkler03AA}
Winkler, C., et al., 2003, A\&A, 411, L1
\end{thebibliography}
\end{document}